\documentstyle[sprocl,epsfig]{article}

\newcommand{\inttp}{\int \frac{d^3p}{(2 \pi)^3}}

\newcommand{\ihsp}{\hspace*{\fill} }

\newcommand{\case}[2]{\mbox{\small $\displaystyle \frac{#1}{#2}$}}

\newcommand{\bcn}{\begin{center}}
\newcommand{\beq}{\begin{equation}}
\newcommand{\beqar}{\begin{eqnarray}}

\newcommand{\ecn}{\end{center}}
\newcommand{\eeq}{\end{equation}}
\newcommand{\eeqar}{\end{eqnarray}}

\newcommand{\Eq}[1]{Eq.~(\ref{#1})}
\newcommand{\Eqs}[1]{Eqs.~(\ref{#1})}
\newcommand{\Fig}[1]{Fig.~\ref{#1}}

\newcommand{\Table}[1]{Table~\ref{#1}}
\newcommand{\sect}[1]{\section{ #1} }
 
\def\be{\begin{equation}}
\def\ee{\end{equation}}
\def\bea{\begin{eqnarray}}
\def\eea{\end{eqnarray}}

\begin{document}

\parbox{108mm}{ 
\footnotesize
Contribution to the Workshop on Future 
Directions in Quark Nuclear Physics, March 9-20, 1998, Adelaide. 
Proceedings to be published. \hspace*{\fill}
\parbox[t]{25mm}{KSUCNR-103-98 \\  nucl-th/9808029 \\ }  }
\normalsize
\title{INSIDE MESONS: COUPLING CONSTANTS AND FORM FACTORS}

\author{PETER C. TANDY}

\address{Center for Nuclear Research, Department of Physics, 
Kent, OH, USA 44242\\E-mail: tandy@cnr2.kent.edu} 


\maketitle\abstracts{ We illustrate the progress of covariant QCD
phenomenology for the description of meson coupling constants and form
factors.  As examples, we discuss the $\rho\pi\pi$ and $\gamma \pi\rho$
interactions, the $\rho$ contribution to the pion charge radius, and
the $\rho NN$ and $\omega NN$ vector and tensor coupling constants and 
form factors. }

\section{QCD Modeling of Mesons and Dressed Quarks}

 In this work the dressed quark
propagators and approximation schemes are guided by the  Dyson-Schwinger 
equation (DSE) approach\cite{DSErev} to non-perturbative QCD modeling of 
hadron physics.
This covariant QCD phenomenology has proved to be quite efficient for 
low-mass mesons and their form factors\cite{T97}.
To expedite investigations we make use of a convenient parameterization 
of confining solutions of quark DSEs.  The broad features  are taken
from the solution to a simple DSE model~\cite{BRW92} that is extremely 
infrared dominant, produces a propagator with no mass-shell pole, and 
includes gluon-quark vertex dressing determined by the Ward
identity.  The resulting propagator is an entire function in 
the complex $p^2$-plane describing absolutely confined~\cite{RWK92} 
dressed quarks in the presence of both explicit and dynamical breaking 
of chiral symmetry.  Additional 
strength for the propagator at intermediate space-like momenta is 
necessary to represent solutions of more realistic DSE models. 
 
With the quark propagator written as \mbox{$S(p)=-i\gamma \cdot
p \sigma_V(p^2)+$}\mbox{$\sigma_S(p^2)$}, the following parameterization
for flavor \mbox{$f= u/d, s$} captures the essential 
features~\cite{R96,BRT96}
\beq
\bar\sigma_S^f(x) =  {\cal F}(b_1^f x) \, {\cal F}(b_3^f x)
\left(b_0^f + b_2^f {\cal F}(\Lambda x)\right) + 2 \bar{m}_f
{\cal F}\left(2(x+\bar{m}_f^2)\right)~ ,
\label{ssb}
\eeq
and
\beq
\bar\sigma_V^f(x) = \frac{2(x + \bar{m}_f^2)
   - 1 + e^{-2(x + \bar{m}_f^2)}}{2(x + \bar{m}_f^2)^2}~.
\label{svb}
\eeq
Here \mbox{${\cal F}(x)=(1-e^{-x})/x$},  \mbox{$x=p^2/\lambda^2$}, 
\mbox{$\bar{\sigma}_V=$}
\mbox{$\lambda^2 \sigma_V$}, \mbox{$\bar{\sigma}_S=$}
\mbox{$\lambda \sigma_S$} with $\lambda$ being the mass scale.  
Also \mbox{$\bar{m}_f = m_f/ \lambda$}, and $\Lambda=10^{-4}$ is not
a free parameter.   
The five parameters \mbox{$\bar m_u,b_0^u,\ldots$}\mbox{$,b_3^u$}
provide a good description of the pion observables:
$f_\pi$; $m_\pi$; $\langle\bar q q\rangle$; $r_\pi$; the $\pi$-$\pi$ 
scattering lengths; and the electromagnetic pion form factor. 
Kaon observables are also fit by making minimal changes to obtain 
the $s$ sector quark parameters.~\cite{BRT96}

The general form of the pion Bethe-Salpeter (BS) amplitude is
\begin{eqnarray}
\Gamma_\pi^j(k;P) & = &  \tau^j \gamma_5 \left[ i E_\pi(k;P) + 
\gamma\cdot P F_\pi(k;P) \rule{0mm}{5mm}\right. \nonumber \\
& & \left. \rule{0mm}{5mm}+ \gamma\cdot k \,k \cdot P\, G_\pi(k;P) 
+ \sigma_{\mu\nu}\,k_\mu P_\nu \,H_\pi(k;P) \right]\, ,
\label{Gammapi}
\end{eqnarray}
and the first three terms are significant in realistic model 
solutions~\cite{MR97} and are necessary to satisfy the axial
Ward identity.~\cite{MRT98}  For convenience, we employ  approximate
$\pi$ BS amplitudes such as those obtained from a rank-2 separable 
ansatz~\cite{sep97} for the ladder/rainbow kernel of the DSE and BSE.  
They preserve Goldstone's theorem and should be adequate for infrared
integrated quantities.   Parameters are fit to $m_{\pi/K}$ and 
$f_{\pi/K}$. The resulting $\pi$ BS amplitude is
\beq
\Gamma_\pi (k,Q) = i\gamma_5 f(k^2)\, \lambda_1\, -
      \gamma_5 \,\gamma\cdot Q f(k^2)\, \lambda_2~. 
\label{sep_pi}
\eeq
The $\rho$ amplitude from the same study~\cite{sep97} is
\beq
\Gamma_\nu(k;Q) = k_{\nu}^Tg(k^2)\lambda_1(Q^2) 
+ i\gamma_{\nu}^T f(k^2)\lambda_2(Q^2) 
+ i\gamma_5 \epsilon_{\mu \nu \lambda \rho}
\gamma_{\mu} k_{\lambda} Q_{\rho}  g(k^2)\lambda_3(Q^2) .
\label{vamp}
\eeq
The BS amplitudes are normalized in the canonical way. 
\begin{figure}[ht]
\ihsp \epsfig{figure=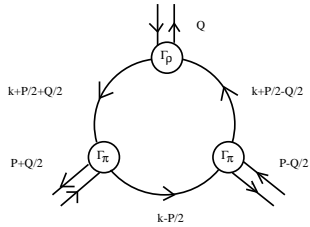,height=4.0cm} \ihsp
\caption{Diagram for the $\rho\pi\pi$ calculation.
\label{diag:rpp} } 
\end{figure}

\sect{The $\rho\pi\pi$ Interaction}
\label{sect_rpp}

\begin{figure}[ht]
\begin{minipage}[t]{58mm}
\ihsp \centering{\
\epsfig{figure=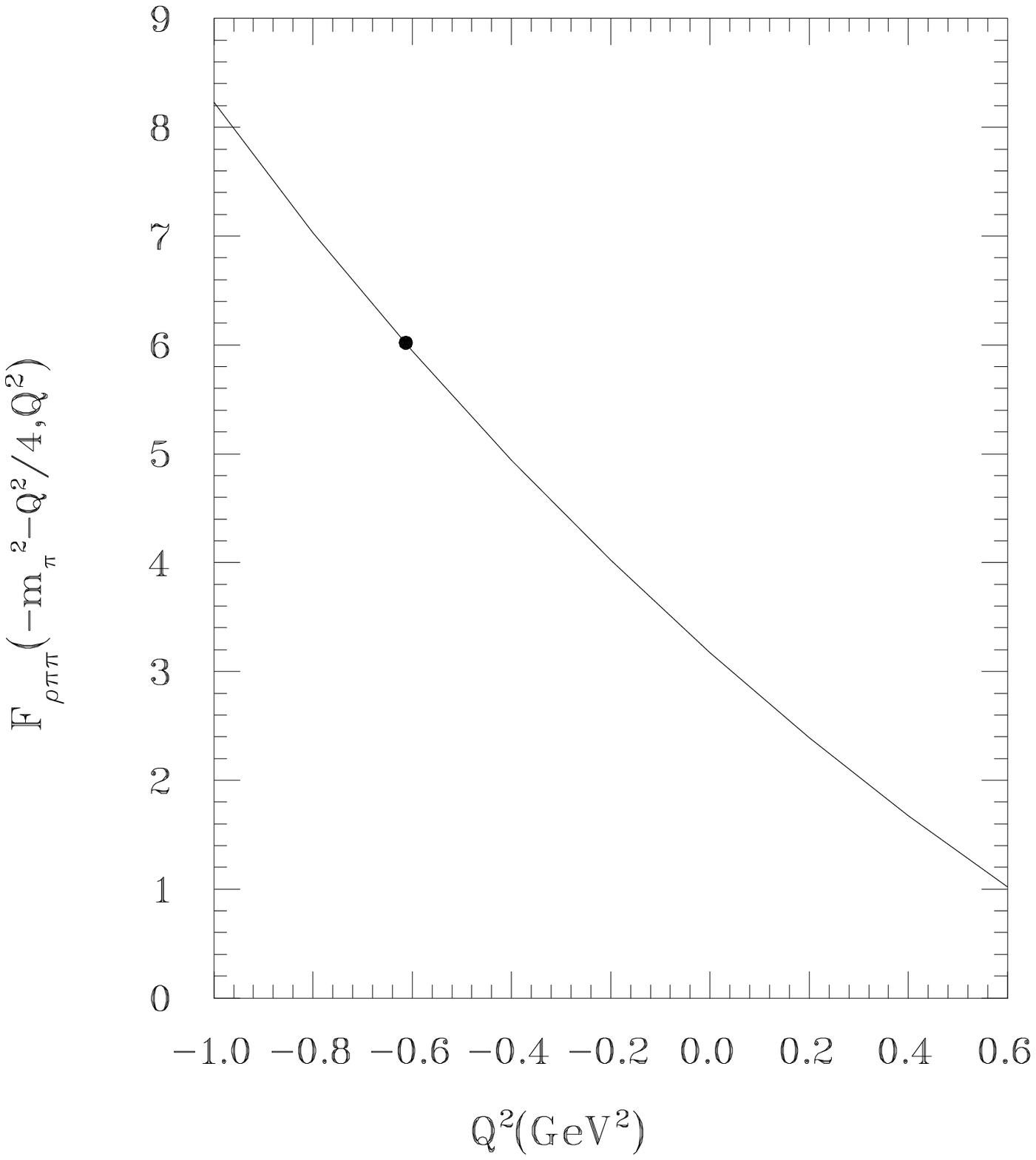,angle=0,height=5.5cm} } \ihsp
\caption{The $\rho\pi\pi$ form factor versus $\rho$ momentum.  The
$\rho$ mass-shell is marked by the dot. \label{rpp} } 
\end{minipage} \hfill \begin{minipage}[t]{58mm}
\ihsp \centering{\epsfig{figure=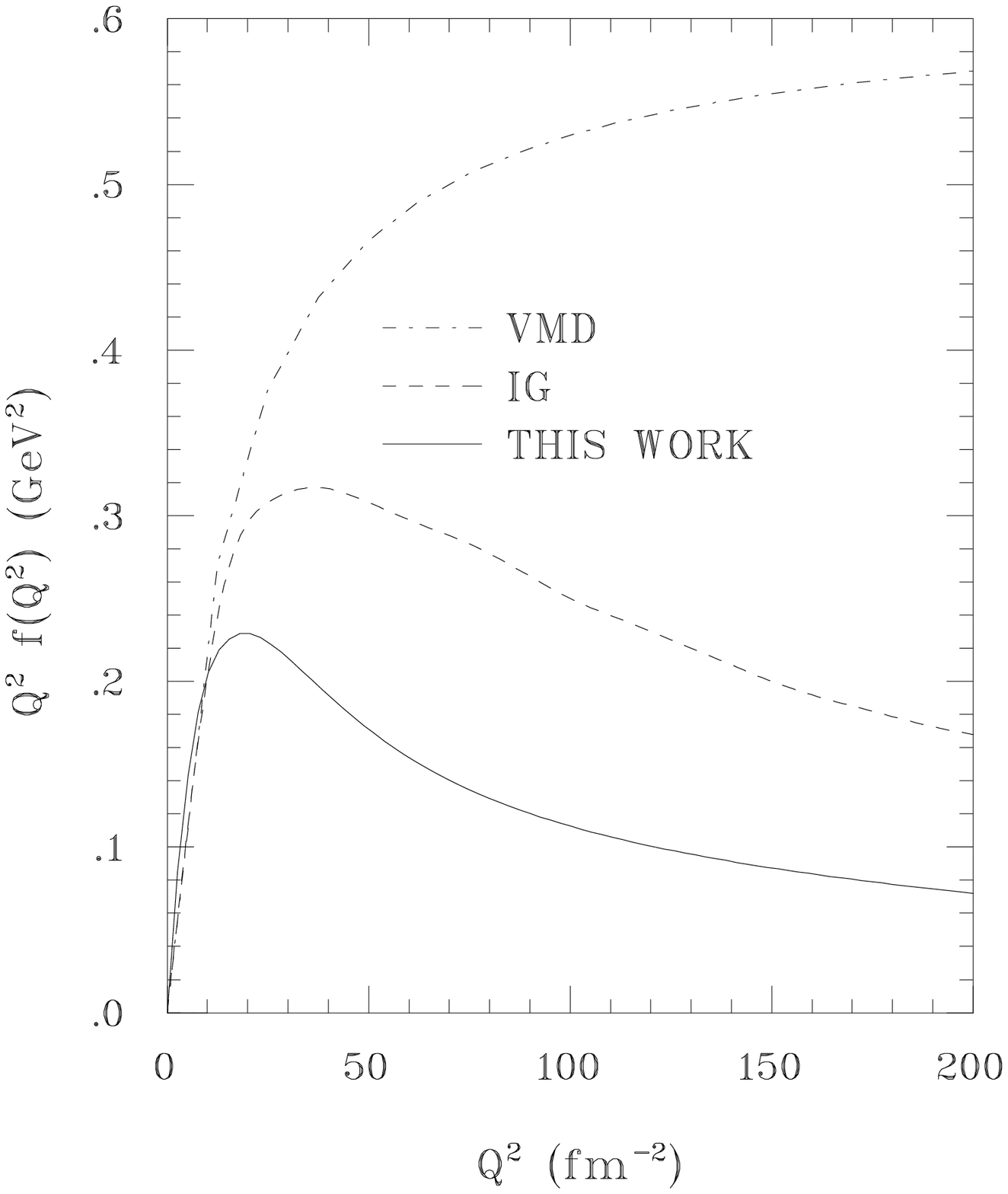,height=5.5cm} } \ihsp
\caption{The $\gamma\pi\rho$ form factor versus photon momentum.
\label{rpg} } 
\end{minipage} 
\end{figure}

Previous attempts to explain the $\rho\pi\pi$ coupling constant in terms
of a covariant quark-gluon phenomenology for the intrinsic properties
of $\rho$ and $\pi$ employed only $\gamma_\mu$ and $\gamma_5$ covariants
for the respective BS amplitudes.~\cite{PRC87bHRM92,MT97}  
It has since been demonstrated for a number of infra-red sensitive 
quantities such as $m_\pi$ and $f_\pi$, that the pseudovector terms
in the pion BS amplitude are responsible for corrections
in the 20-30\% range.~\cite{MR97,MRT98,sep97}   Here we reexamine the
\mbox{$\rho\pi\pi$} interaction with more general BS
amplitudes, such as
\beq
\vec{\Gamma}_\pi (k;Q)= \vec{\tau} \sum_i {\cal K}_\pi (i) 
\Gamma_\pi^i (k;Q)~, 
\label{genpibs}
\eeq
for the $\pi$. 
Here $Q$ is the $\pi$ momentum, $k$ is the relative $\bar q q$ 
momentum, ${\cal K}_\pi(i)$ is the i$^{th}$ covariant
constructed from gamma matrices and momenta, and $\Gamma_\pi^i(k;Q)$
is the corresponding totally scalar amplitude.   We use a parallel 
notation for the $\rho$ amplitude. 
The first term in a skeleton graph expansion of the $\rho\pi\pi$ 
vertex~\cite{MT97} yields
\beq
\Lambda_\mu(P,Q)=\int \frac{d^4k}{(2\pi)^4}
\Gamma_\rho^i(k^\prime;Q^\prime) 
\Gamma_\pi^j(k^{\prime\prime};Q^{\prime\prime}) 
\Gamma_\pi^k(k^{\prime\prime\prime};Q^{\prime\prime\prime}) 
T_\mu^{ijk}~,
\label{vint}
\eeq
where the required discrete traces are
\beq
T_\mu^{ijk}(k,P,Q)= 2 N_c {\rm tr}_s \left[
S(q^\prime) {\cal K}^\prime_\mu(i)
S(q^{\prime\prime}) {\cal K}_\pi^{\prime\prime}(j)
S(q^{\prime\prime\prime}) 
{\cal K}_\pi^{\prime\prime\prime}(k) \right].
\eeq
Summation over the labels (ijk) for the various meson covariants
is implied.   With reference to Fig.~\ref{diag:rpp},
the systematic  notation for momenta is:  
the $\rho$ vertex is characterized by
single prime quantities (\mbox{$k^\prime; Q^\prime \equiv Q$}), the 
pion vertex bringing $P-Q/2$
into the loop is double-prime, the other pion vertex bringing
$-(P+Q/2)$ into the loop is triple-prime;  the outgoing 
quark momentum from the prime vertex is $q^\prime$, similar for the
other vertices.  Thus \mbox{$k^\prime = $}\mbox{$(q^\prime -
q^{\prime\prime})/2$}, etc.   With both pions on the mass-shell,
$P\cdot Q=0$ and \mbox{$P^2=$}\mbox{$-m_\pi^2-\frac{Q^2}{4}$}.  In this
case symmetries require the form
\mbox{$\Lambda_\mu(P,Q)=$}\mbox{$-P_\mu  F_{\rho\pi\pi}(Q^2)$} and
the coupling constant is 
\mbox{$ g_{\rho\pi\pi}=$}\mbox{$ F_{\rho\pi\pi}(Q^2=-m_\rho^2)$}.  
\begin{table}[ht]
\caption{$g_{\rho\pi\pi}$ calculation and contributions from meson 
covariants. \label{tab:rpp} }
\vspace{0.2cm}
\begin{center}
\footnotesize
\begin{tabular}{|c|}\hline
 $g_{\rho\pi\pi}= 6.28$  [expt 6.05] \\
\end{tabular}\\
\begin{tabular}{|cl||cl|} \hline
  $\pi$ Covariants  &    & $\rho$ Covariants & \\ \hline
 $\gamma_5$ & 171\%  & $\gamma_\mu$ & 94.5\%   \\ 
 $\gamma_5 \gamma \cdot Q$ & -71\% & $\gamma_5 \epsilon_\mu \; 
                                           \gamma k Q$ & 5.5\% \\ 
            &      &     $k_\mu$   &     0.01\%  \\ \hline
\end{tabular}
\end{center}
\end{table}
With the separable model BS amplitudes of \Eqs{sep_pi} and (\ref{vamp}),
the prediction for $g_{\rho\pi\pi}$, given in \Table{tab:rpp}, compares
favorably with the empirical value associated with the 
$\rho\rightarrow\pi\pi$ decay width. Truncation to the dominant 
$\rho$ amplitude is found
to only make a 5\% error.  However the sub-dominant pion component 
(pseudovector \mbox{$\gamma_5 \gamma \cdot Q$}) enters quadratically here
and makes a major contribution (-71\%). 
The calculated form factor $F_{\rho\pi\pi}(Q^2)$ is shown in 
Fig.~\ref{rpp}. Of course the detailed shape of a
calculated form factor depends upon the definition of the composite meson 
propagator.  It is always subject to 
a field redefinition.  We have consistently renormalized so that all
momentum dependence other than that of the standard point meson 
propagator is allocated to the vertex function.~\cite{T97}

\sect{ The $\gamma\pi\rho$ Interaction}
\label{sect_gpr}

The isoscalar $\gamma^\ast \pi\rho$ meson-exchange current contributes 
significantly to electron scattering from light nuclei. 
Our understanding of the deuteron EM structure functions
for $Q^2 \approx 2-6~{\rm GeV}^2$ requires knowledge of this form 
factor.~\cite{VODG95}
Available studies of the vertex function 
$\Lambda_{\mu\nu}(P,Q)$ use just a $\gamma_5$ covariant for the
$\pi$ (with amplitude $E_\pi$) and a $\gamma_\mu^T$ covariant for 
the $\rho$ (with amplitude $V_\rho$).   At the quark 
loop level, the expression is
\beqar
\Lambda_{\mu\nu}(P,Q)&=&\case{e}{3}\int \frac{d^4k}{(2\pi)^4}
E_\pi(k+\case{Q}{4};-P-\case{Q}{2})
V_\rho(k-\case{Q}{4};P-\case{Q}{2}) \nonumber \\
&\times& 2 N_c {\rm tr}_s \big[S(k_+ -\case{Q}{2})\Gamma_\nu(k_+;Q)
S(k_+ +\case{Q}{2})i\gamma_5 S(k_-)i\gamma_\mu^T \big] ~,
\label{lam}
\eeqar
where $\Gamma_\nu(k_+;Q)$ is the photon vertex. Here 
\mbox{$k_\pm = k \pm \frac{P}{2}$}.  The bare photon vertex 
$- i \gamma_\mu $ is clearly inadequate for dynamically dressed quarks
because it violates the Ward-Takahashi identity (WTI).  
We employ the Ball-Chiu~\cite{BC80} ansatz  for $\Gamma_\nu$ because 
it obeys the relevant symmetries and is conveniently 
determined completely in terms of the quark propagator.  
Then the WTI gives \mbox{$Q_\nu\;\Lambda_{\mu\nu}=0$}; the 
$\gamma\pi\rho$ current is conserved.
The  vertex function has the general form
\beq
\Lambda _{\mu \nu }(P,Q) = -i\frac{e}{m_{\rho}}\; \epsilon
_{\mu \nu \alpha \beta } \; P_{\alpha }Q_{\beta }~g_{\rho \pi \gamma}~
f(Q^2,P^2,P\cdot Q) ~. \label{gprgen}
\eeq

The available calculation uses \mbox{$E_\pi(q;P) = B(q^2,m)/f_\pi$}, 
where $B$ is the quark scalar self-energy,  and also
\mbox{$V_\rho(p^2)\propto e^{-p^2/a^2}$}.  The norm of $V_\rho$ 
is set in the canonical way.  The range $a$ is adjusted to reproduce
$g_{\rho\pi\pi}^{\rm expt}=6.05$. This has proved to be a good
phenomenological basis predicting other $\rho$ 
processes.~\cite{MTRC94,PL96}  
The resulting prediction $g_{\gamma \pi \rho}=0.5$ agrees well with the
empirical value $g_{\gamma \pi \rho}^{\rm expt}=0.54\pm 0.03$ from $\rho$
decay.  The $\gamma^\ast \pi\rho$ form factor weighted by $Q^2$ is 
shown in \Fig{rpg}.   The result is much softer than either the
vector meson dominance (VDM) prediction or a quark loop without
momentum-dependent dressing.  The available data for elastic 
EM deuteron form factors $A(Q^2)$ and $B(Q^2)$ in the range 
$2-6~{\rm GeV}^2$ ($50-150~{\rm fm}^{-2}$)  has been shown to 
strongly favor the present result over a variety of other 
approaches.~\cite{VODG95}  The  $\gamma \pi \rho$ mechanism for 
vector meson electroproduction has been
treated in a closely related way.~\cite{P98}

\sect{Vector Meson Role in the Pion Form Factor}
\label{sect_vmdpiff} 

An important question is the size of intermediate state meson  mechanisms 
or meson loop corrections.   When the composite and extended 
structure of the meson modes is accounted for, the distributed vertex 
functions tend to leave meson loops less of a role than for models 
or effective field theories built on point coupling.  Studies 
of the $\rho-\omega$ mass difference with $\bar q q$ composite 
pion loop dressing~\cite{PRC87bHRM92,MT97} confirm this.
Here we outline an analysis of the role for the $\rho$ in the 
space-like pion charge form factor.   The dressed photon-quark 
vertex $\Gamma_\nu(q;Q)$ can be separated (non-uniquely) into a 
$\rho$ pole or resonant piece (which is 
transverse) and a background or non-resonant piece (which is both 
longitudinal and transverse).  That is,~\cite{T97} 
\beq
\Gamma_\nu(q;Q) = \Gamma_\nu^{nr}(q;Q) - \Gamma^\rho_\mu(q;Q) 
\; \frac{T_{\mu \sigma}(Q)}{Q^2 + m_\rho^2(Q^2)}
\; \Pi^{\rho \gamma}_{\sigma \nu}(Q)  .
\label{polevertex}
\eeq
The $\rho\gamma$ polarization tensor is given, in abbreviated notation, by
\beq
\Pi^{\rho \gamma}_{\sigma \nu}(Q) = {\rm Tr} \; [ \bar 
\Gamma_\sigma^\rho(-Q) \; S \; \Gamma_\nu(Q) \; S ]  .
\label{grmixing}
\eeq
{}From \Eq{polevertex}, the pion charge form factor takes the form
\beq
{}F_\pi(Q^2) = F_\pi^{GIA}(Q^2) + \frac{ F_{\rho\pi\pi}(Q^2) 
                   \; \Pi^{\rho \gamma}_T(Q^2) } {Q^2 + m_\rho^2(Q^2)} ~.
\label{piffrho}
\eeq
With the Ball-Chiu Ansatz used for $\Gamma_\nu^{nr}$, the resulting 
generalized impulse approximation (GIA) has been found to be 
phenomenologically successful in describing the spacelike $F_\pi(Q^2)$.  
A persistent result  is that $85-90$\% of the charge radius 
is naturally explained.~\cite{R96}   The $\rho$ resonant term is 
obviously necessary near the timelike pole.   However 
the relative contribution of the two terms in the spacelike 
region, depends upon how the spectral strength of the underlying 
$q \bar q$ scattering kernel is divided into pole and background.   
While the individual terms depend upon choice of interpolating field, 
the summed contribution to an S-matrix element should not.

In the  limit of a point coupling model with structureless hadrons, 
the elements
of \Eq{piffrho} become \mbox{$F_\pi^{GIA}(Q^2) \rightarrow 1$}, 
\mbox{$F_{\rho \pi\pi}(Q^2) \rightarrow g_{\rho \pi\pi}$}, and 
\mbox{$ \Pi^{\rho \gamma}_T(Q^2) \rightarrow $} \mbox{$-Q^2/g_V$}.  
This produces the VMD empirical form
\beq
{}F_\pi(Q^2) = 1 - \frac{ g_{\rho\pi\pi} \; Q^2} { g_V \, (Q^2 + m_\rho^2) } ~.
\label{piffvmd}
\eeq
This implies 
\mbox{$r_\pi^2 \sim 6  g_{\rho \pi\pi}/(m_\rho^2 g_V) $}, and with 
universal vector coupling, produces 
\mbox{$r_\pi^2 \sim 0.4~{\rm fm}^2$}, most of the experimental
value $0.44~{\rm fm}^2$. 
\begin{figure}[ht]
\vspace*{1.0cm}
\ihsp \epsfig{figure=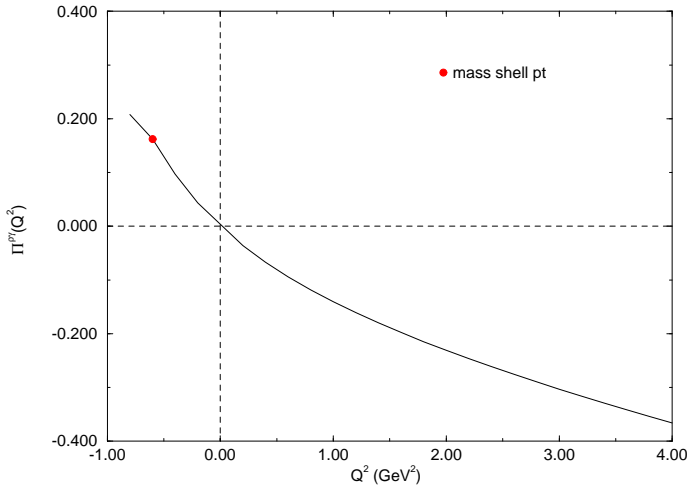,height=5.5cm} \ihsp
\caption{The transverse $\rho \gamma$ mixed self-energy amplitude.  The 
$\rho$ mass-shell is indicated.  The zero at \mbox{$Q^2=0$} preserves
the massless photon. \label{fig:rg} } 
\end{figure}
Given 85\% of $r_\pi$ from the GIA, is the $\rho$ contribution  small 
enough in the present QCD-modeling approach?   Electromagnetic gauge 
invariance requires the $\rho$-$\gamma$ mixing amplitude 
\mbox{$\Pi^{\rho \gamma}_T(Q^2)$}, and hence the pole term of 
$F_\pi(Q^2)$ in 
\Eq{piffrho}, to vanish at $Q^2=0$.   Asymptotic freedom ensures 
\mbox{$\Pi^{\rho \gamma}_T(Q^2)$} vanishes at large spacelike-$Q^2$.    
We use the representations
\mbox{$F_{\rho \pi\pi}(Q^2) =  g_{\rho \pi\pi} f_{\rho \pi\pi}(Q^2)$} and
\mbox{$\Pi^{\rho \gamma}_T(Q^2) = -Q^2 \hat{f}_{\rho \gamma}(Q^2)/g_V $}, 
where $f$ and $\hat{f}$ depend on meson substructure dynamics.   The 
$\rho$ contribution to $r_\pi$ from \Eq{piffrho} is
\beq
(r_\pi^{pole})^2 = r_\pi^2 - (r_\pi^{GIA})^2 = 1.2 \;
 f_{\rho \pi\pi}(0)  \hat{f}_{\rho \gamma}(0) \; \frac{6}{m_\rho^2}  ~,
\label{rpipole}
\eeq
where we have used the empirical result 
\mbox{$ g_{\rho \pi\pi}/g_V \sim 1.2$}. 
{}From \Fig{rpp}, \mbox{$f_{\rho \pi\pi}(0) \approx$} \mbox{$ 0.5$} and
our calculation of $\Pi^{\rho \gamma}_T(Q^2)$ from \Eq{grmixing} gives
\mbox{$ \hat{f}_{\rho \gamma}(0) = 0.65$}.    This yields 
\mbox{$(r_\pi^{pole})^2 =$} \mbox{$ 0.16~{\rm fm}^2$} in contrast to the 
VMD result of  $0.4~{\rm fm}^2$.  With  
\mbox{$(r_\pi^{GIA})^2 =$} \mbox{$ 0.31~{\rm fm}^2$}, the present 
approach gives a total of $0.47~{\rm fm}^2$ compared to
the experimental value ($0.44~{\rm fm}^2$).  This is obviously an
overestimate leaving no room for
the pion loop contribution of the expected~\cite{ABR95} size.
However, the main point is that a $\rho$ contribution to the pion charge
radius is a model-dependent quantity and a value much smaller than that
from the simple VMD assumption is consistent with the present staus of
DSE-based QCD modeling of the pion.

\section{$\rho NN$ and $\omega NN$ Couplings}

The dynamical content of the simple vector meson BS amplitude of \Eq{vamp} 
has been found~\cite{TQB98} to produce $\rho NN$ and $\omega NN$ couplings
consistent with empirical values deduced from boson exchange model fits to
NN data.  We use a mean field chiral 
quark-meson model of the nucleon~\cite{FT92,B98} for which the internal 
chiral meson modes are generated as $\bar{q}q$ correlations.  
In a Euclidean metric, the relevant nucleon vector current 
\beq
J_N^\mu(-Q)=\case{1}{Z_N}\langle N|\inttp \; 
\bar q(p+\case{Q}{2})\;\Gamma_\nu(p;Q)\; q(p-\case{Q}{2})|N \rangle
\label{j}
\eeq
where $q$ is the quark field, $Q$ is the meson momentum, $|N \rangle$ is 
the static mean field nucleon state and 
$\Gamma_\nu$ is the BS amplitude.  The nucleon valence quark wave 
function renormalization constant $Z_N$ arises from the dynamical  
nature of the quark self-energy.~\cite{FT92} 

At the meson mass-shell, 
$\Gamma_\nu$ is normalized in the canonical way such
that it is the residue of the vector $\bar q q$ propagator there.  
However for the $NN$ interaction, spacelike $Q^2$ is needed.  An 
appropriate strategy there is to use the BS eigenvalue problem
\mbox{${\cal K}_L(Q) \Gamma_\nu(Q) = \lambda(Q^2)\Gamma_\nu(Q)$} 
where ${\cal K}_L$ is the BSE kernel.  One can express the approach to the
mass shell as \mbox{$\lambda(Q^2) = 1 -$}\mbox{$(Q^2 + M_V^2)$}
\mbox{$ Z_V^{-1}(Q^2)$}
where $Z_V$ is unity at the mass shell.  The consistent definition of 
propagator for vector $\bar q q $ correlations in this approach is the 
$\bar q q$ scattering operator or T-matrix \mbox{$T=D-{\cal K}_L T$} 
where $D$ represents the gluon $2$-point function. For general $Q^2$, 
we normalize $\Gamma_\nu$ so that the propagator for vector 
$\bar q q$ correlations has the mode expansion form
\beq
{\cal D}_{\mu \nu}^T(q',q;Q) =\sum_n \; \frac{\Gamma_\mu^T(q';Q;n) \otimes 
\bar{\Gamma}_\nu^T(q;-Q;n)} {(Q^2+M_n^2)} \; \, \rightarrow \; \, 
\frac{i\gamma_\mu^T g(Q^2) \otimes i\gamma_\nu^T g(Q^2)} {Q^2+M^2} ~.
\label{exp}
\eeq 
All momentum dependence except for the standard point particle denominator
has been moved into the generalized BS amplitudes.
\Eq{exp} also shows the point-coupling limit appropriate to the 
Nambu--Jona-Lasinio model~\cite{NJL} where there is no dependence 
on $\bar q q$ relative momentum.  

The form factors $F_1$ and $F_2$ are identified from
recasting the results from Eq.~(\ref{j}) into the form
\mbox{$J_N^\mu(-Q)=$} \mbox{$\bar u(\vec p~')[\;i\gamma_\mu F_1(Q^2)+$}
\mbox{$i\frac{\sigma_{\mu\nu}}{2M}Q_\nu F_2(Q^2)\;]u(\vec p)$}.
\begin{figure}[htb]
\begin{minipage}[t]{58mm}
\ihsp \centering{\epsfig{figure=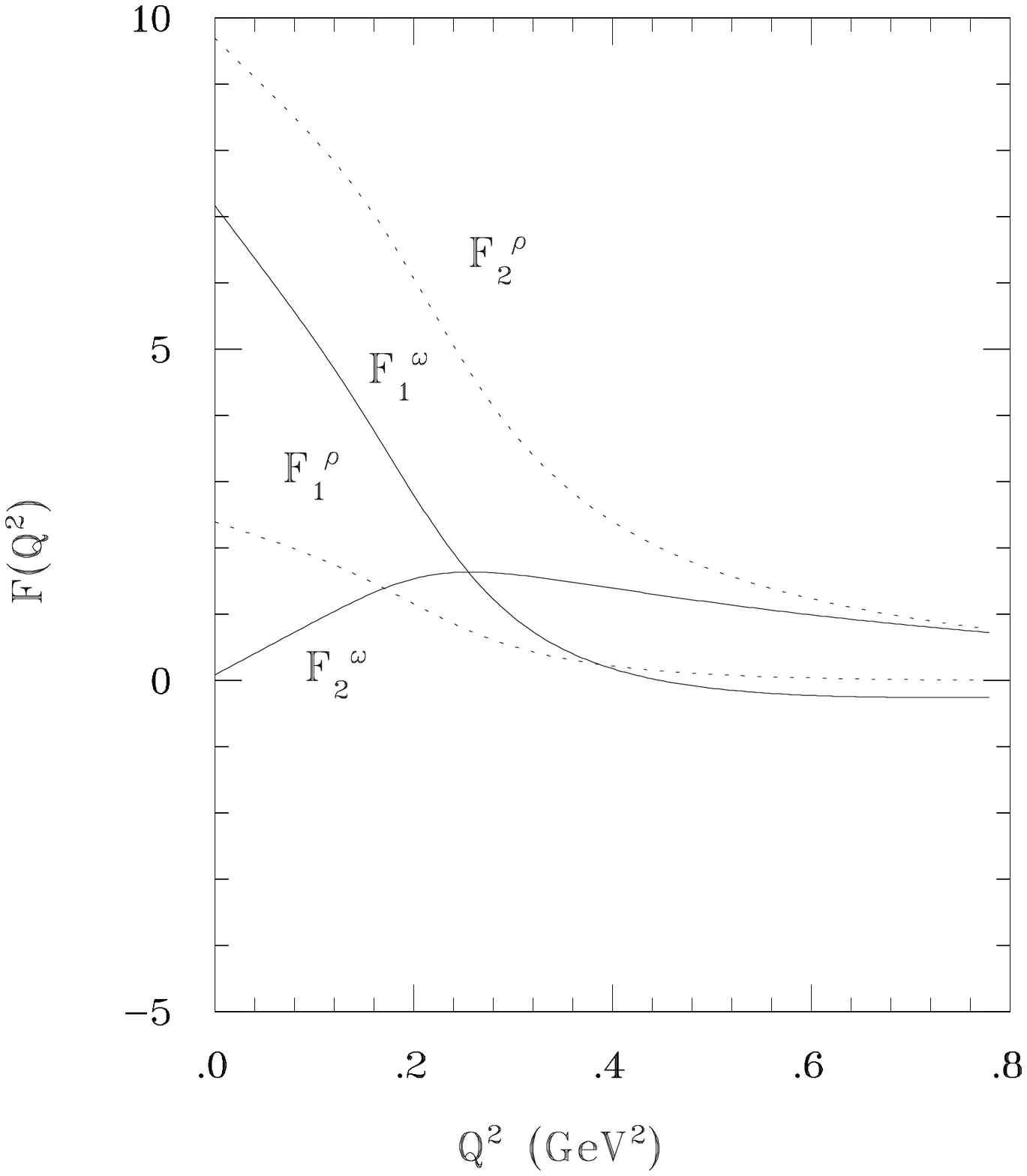,height=5.5cm} } \ihsp
\caption{Vector and tensor form factors using the three covariant form 
of the  BS amplitude.} 
\label{fig:3}
\end{minipage} \hfill \begin{minipage}[t]{58mm}
\ihsp \centering{\epsfig{figure=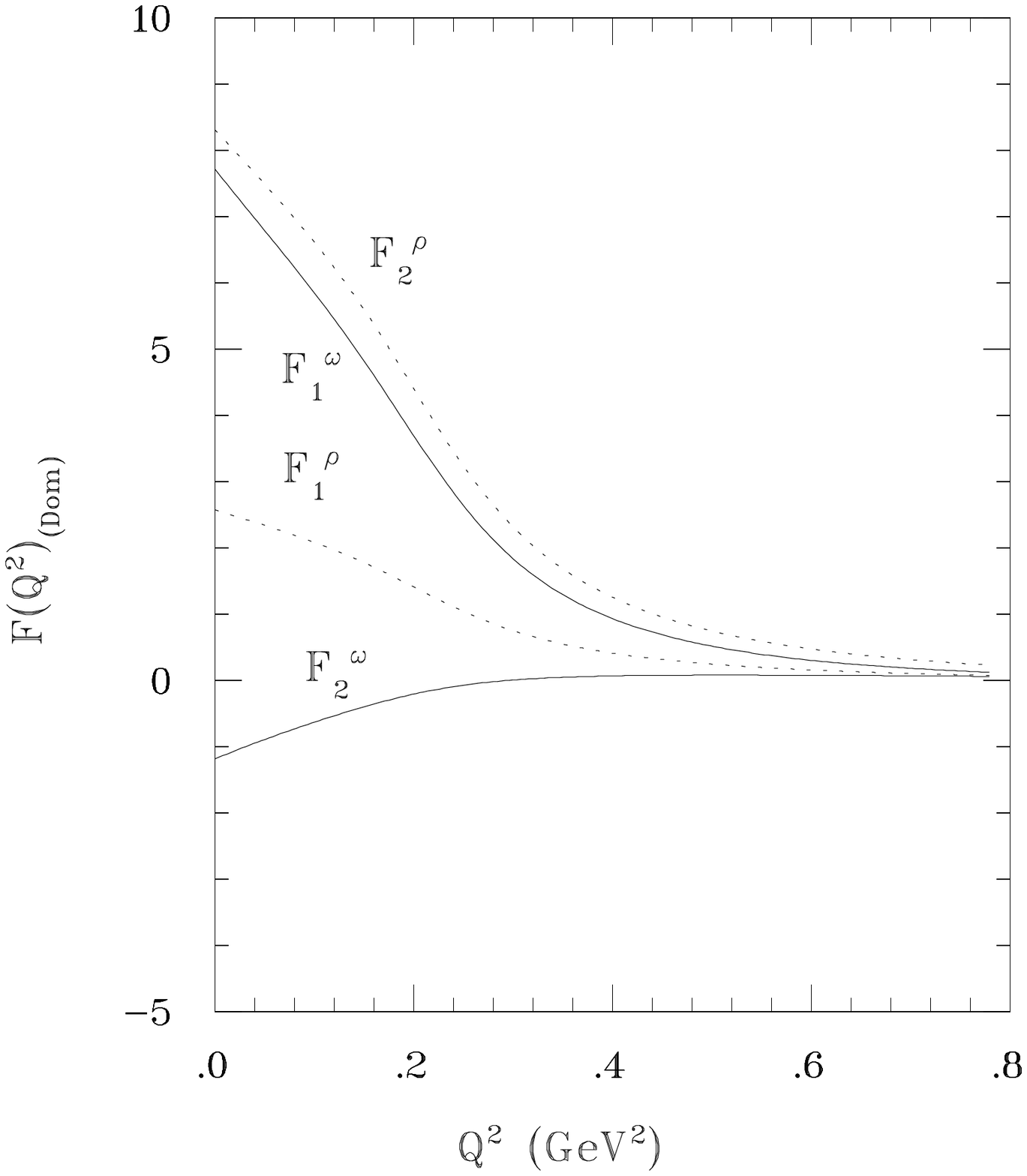,height=5.5cm} } \ihsp
\caption{Only the dominant BS covariant amplitude is used here.}
\label{fig:4}
\end{minipage} 
\end{figure}
In Figs.~\ref{fig:3} and \ref{fig:4} we show the results~\cite{TQB98} for the 
$\rho N N$ and $\omega N N$ form
factors obtained from the full BS amplitude of Eq.~(\ref{vamp}).  
The coupling strengths \mbox{$F_{1/2}(Q^2=0)$}, which are the relevant measures for
the NN interaction,  are given in the Table.  
Coupling constants from extrapolation to the mass-shell via a typical boson 
exchange 
monopole form factor with range \mbox{$\Lambda=1.5~{\rm GeV}$} are shown in 
parenthesis.  The momentum dependence and strength of the vector meson 
BS amplitudes are seen to be consistent with the qualitative features of 
the Bonn~\cite{BONN} boson-exchange  NN model. 
A more recent value for the poorly-determined
$g_{\omega NN}$ is $7-10.5$ from analysis~\cite{SL96} of pion photoproduction on the nucleon.
\begin{table}[hbt]
\caption{Coupling constants and coupling strengths for $\rho NN$ and $\omega NN$ \label{tab:ccs} }
\vspace{0.2cm}
\begin{center}
\footnotesize

\begin{tabular}{|c|c|c|}
\hline
    & Prediction  & Empirical-OBE~\cite{BONN}  \\
\hline
 $F_1^\omega(0)$ ($g_{\omega NN}$)   & 7.2 (9.9)     & 11.7  (16) \\
 $F_2^\omega(0)$ ($f_{\omega NN}$)   & 0.08 (0.11)    & 0  (0) \\
 $F_1^\rho(0)$  ($g_{\rho NN}$)    & 2.4 (3.3)      & 2.6  (3.5) \\
 $F_2^\rho(0)$  ($f_{\rho NN}$)    & 9.7 (13.3)      & 16.1 ( 22) \\
 $\kappa_{\rho}=\frac{f_{\rho NN}}{g_{\rho NN}}  $  &4.0 & 6.1 \\
 $\kappa_{\omega}=\frac{f_{\omega NN}}{g_{\omega NN}}  $ &0.01  & 0 \\
 $\frac{\kappa_\rho}{\kappa_\gamma}  $   &1.09  & 1.78 \\
\hline
\end{tabular}
\end{center}
\end{table}
Note that since this nucleon model overestimates the magnitude of the
magnetic moments by \mbox{$\sim 15\%$},~\cite{B98} one may expect at 
least this amount of uncertainty here also.

\section{Summary}
 
Since the parameters in this approach have been previously fixed through the 
requirement that soft chiral quantities such as $m_{\pi/K}$, $f_{\pi/K}$ and charge
radii $r_{\pi/K}$ be reproduced, the meson couplings discussed here
have been produced without adjusting parameters.   The results imply 
that this present approach to modeling QCD for low-energy hadron physics 
can capture the dominant infrared physics.  We expect that the large 
momentum behavior of form factors such as $\gamma \pi\pi$ and 
$\pi \gamma \gamma $ will require attention to more detailed aspects
of the dynamics.
  
\section*{Acknowledgments}
Thanks are due to A. W. Thomas for generous hospitality and the 
organization 
of a fine working environment for the Workshop.  This work is partly
supported  by the US National Science Foundation under grant 
No. PHY97-22429.

\end{document}